%% file: impurity_model.tex
\documentclass[11pt]{article}

\usepackage{hepth}
\usepackage{array}
\usepackage{hyperref}
\usepackage{graphicx}

\begin{document}
\renewcommand{\topfraction}{0.7} 
\renewcommand{\bottomfraction}{0.7} 

\input{defs}

\input{title}

\input{intro}

\input{impurity}

\input{concs}

\section*{Acknowledgments}
I am grateful to I.~Bakas, A.~Cappelli, D.~Minic, T.~Petkou, K.~Sfetsos and S.~Sachdev for stimulating discussions, and to G.~Grignani for suggesting some references. I would also like to thank the Galileo Galilei Institute for Theoretical Physics for the hospitality and the INFN for partial support during the completion of this work.

\begin{appendix}
\input{wilson_polyakov}

\end{appendix}

\begin{singlespace}
\bibliographystyle{JHEP}
\bibliography{impurity_model}
\end{singlespace}
\end{document}

%% file: defs.tex
%  definitions

% i.e. and e.g.
\newcommand{\ie}{i.e.,\ }
\newcommand{\eg}{e.g.,\ }

% const
\newcommand{\const}{\operatorname{const.}} 

% sgn
\providecommand{\sgn}{\operatorname{sgn}} 

% diag
\providecommand{\diag}{\operatorname{diag}}

% differentials (roman d)
\newcommand{\rmd}{\,\mathrm{d}}

% trace
\newcommand{\Tr}{\operatorname{tr}}

% identity matrix
\newcommand{\idmat}{\mathbb{I}}

% Re and Im
\newcommand{\re}{\operatorname{Re}}
\newcommand{\im}{\operatorname{Im}}

% base of exponentials (roman e), with argument 
\newcommand{\e}[1]{\operatorname{e}^{#1}}

% roman function names
\newcommand{\hypF}[1]{\operatorname{F}\left(#1\right)}
\newcommand{\hypFApp}[1]{\operatorname{F_1}\left(#1\right)}
\newcommand{\B}[1]{\operatorname{B}\left(#1\right)}

% operator
\newcommand{\op}{\mathcal{O}}

% vev
\newcommand{\vev}[1]{\left\langle #1 \right\rangle}

% bold vectors
\newcommand{\bfx}{\mathbf{x}}
\newcommand{\bfy}{\mathbf{y}}
\newcommand{\bfS}{\mathbf{S}}
\newcommand{\bfv}{\mathbf{v}}
\newcommand{\bff}{\mathbf{f}}

% components of x'
\newcommand{\yper}{y_{\perp}}
\newcommand{\ypar}{y_{\parallel}}
\newcommand{\cper}{c_{\perp}}
\newcommand{\cpar}{c_{\parallel}}

% combinatorial factors
\newcommand{\comb}[2]{\begin{pmatrix} #1\\#2\end{pmatrix}} 

% e^\Phi/g_s
\newcommand{\ephi}{\left(\frac{\e{\Phi}}{g_s}\right)}

% black hole radius
\newcommand{\rh}{r_+}

% cut-off
\newcommand{\rc}{r_c}

% S_bound
\newcommand{\Sbulk}{I_{D5}}
\newcommand{\Sb}{I_{\text{bound}}}

% action prefactor
\newcommand{\prefac}{\frac{N}{3\pi^2 \alpha'}}

% Lagrangian
\newcommand{\Lag}{\mathcal{L}}
\newcommand{\Lbulk}{\Lag_{\text{bulk}}}
\newcommand{\Lbound}{\Lag_{\text{boundary}}}

% 2-d induced metric and gauge-field
\newcommand{\G}{\mathcal{G}}
\newcommand{\F}{\mathcal{F}}
\newcommand{\A}{\mathcal{A}_\tau}

% AdS length scale
\newcommand{\lad}{l}

% effective string tension
\newcommand{\Teff}{T_{\text{eff}}}

% derivatives of coordinates
\newcommand{\tdt}{\dot{t}}
\newcommand{\tpr}{t'{}}
\newcommand{\rdt}{\dot{r}}
\newcommand{\rpr}{r'{}}
\newcommand{\xdt}{\dot{\bfx}}
\newcommand{\xpr}{\bfx'{}}
\newcommand{\thdt}{\dot{\theta}}
\newcommand{\thpr}{\theta'{}}

% functions a(r) and b(r)
\newcommand{\ar}{a(r)}
\newcommand{\br}{b(r)}

% Order od magnitude
\newcommand{\Order}{\mathcal{O}}

\newcommand{\barDf}{$\overline{\text{D5}}$}

% units
\newcommand{\unit}[1]{\operatorname{#1}}

% subscripts for disconnected and connected configurations
\newcommand{\Wilson}{\text{Wilson}}
\newcommand{\Dimer}{\text{Dimer}}
\newcommand{\one}{\text{Polyakov}}
\newcommand{\two}{\parallel}
\newcommand{\dimer}{\cup}

% redefined E, F and T and c
\newcommand{\EN}{\mathcal{E}}
\newcommand{\FE}{\mathcal{F}}
\newcommand{\Temp}{\mathcal{T}}
\newcommand{\Entr}{\mathcal{S}}
\newcommand{\spheat}{c_x}

% reduced elasticity
\newcommand{\tkap}{\tilde{\kappa}}

% tilde A
\newcommand{\tA}{\tilde{A}}

% chemical potential
\newcommand{\chp}{\bar{\mu}}

% tilde tau and omega
\newcommand{\ttau}{\tilde{\tau}}
\newcommand{\tom}{\tilde{\omega}}
\newcommand{\tnu}{\tilde{\nu}}

% principle value
\newcommand{\Pval}{\mathcal{P}}

% residue
\newcommand{\res}{\operatorname{Res}}

% spectral asymmetry angle
\newcommand{\vt}{\vartheta}

% Luttinger Ward functional
\newcommand{\LW}{\Phi_{\text{LW}}}
 
% hatted quantities
\newcommand{\hlambda}{\hat{\lambda}}
\newcommand{\hmu}{\hat{\mu}}

% gYM
\newcommand{\gYM}{g_{\text{YM}}}

%% file: title.tex
\preprint{NA-DSF-12/2010}
\title{The Polyakov Loop of Anti-symmetric Representations as a Quantum Impurity Model}

\authors{Wolfgang M{\"u}ck\footnote{e-mail: \texttt{wolfgang.mueck@na.infn.it}}}

\institution{Naples}{Dipartimento di Scienze Fisiche, Universit\`a degli Studi di Napoli "Federico II" \cr and 
INFN, Sezione di Napoli, Via Cintia, 80126 Napoli, Italy}

\abstract{The Polyakov loop of an operator in the anti-symmetric representation in $\mathcal{N}=4$ SYM theory on spacial $\mathbb{R}^3$ is calculated, to leading order in $1/N$ and at large 't~Hooft coupling, by solving the saddle point equations of the corresponding quantum impurity model. Agreement is found with previous results from the supergravity dual, which is given by a D5-brane in an asymptotically AdS$_5 \times S^5$ black brane background. It is shown that the azimuth angle, at which the dual D5-brane wraps the $S^5$, is related to the spectral asymmetry angle in the spectral density associated with the Green's function of the impurity fermions. Much of the calculation also applies to the Polyakov loop on spacial $S^3$ or $H^3$.}

%\PACS{}
\date{January 2011}

\maketitle

\tableofcontents

%% file: intro.tex
\section{Introduction}
\label{intro}

It has been argued recently that holographic techniques can be used to study the strange metal behaviour of correlated electron compounds \cite{Lee:2008xf, Liu:2009dm, Cubrovic:2009ye, Faulkner:2009wj, Denef:2009yy, Hartnoll:2009ns,  Faulkner:2010tq, Faulkner:2010da, Faulkner:2010zz, Larsen:2010jt, Hartnoll:2010gu, Puletti:2010de, Hartnoll:2010xj}. The general idea is to model the CFT corresponding to the quantum critical point holographically and introduce a chemical potential in order to obtain a system with finite density. Typically, this leads to an AdS$_2$ factor in the dual geometry, which turns out to be the key to the unusual properties of the Fermi surface of strange metals. It has also been proposed that holographic strange metals are closely related to Kondo lattice models describing quantum impurities coupled to the CFTs that arise from two-dimensional quantum anti-ferromagnets \cite{Sachdev:2010um, Sachdev:2010uj}. Depending on the value of a certain tuning parameter, the ground state of these anti-ferromagnets can be such that the spins at neighbouring lattice sites pair up into dimers of total spin zero. Holographic dimer models have been considered recently in \cite{Kachru:2009xf}. The first order phase transition that occurs between a ``normal'' and a dimerized phase has been argued to induce a phase transition between the Fermi and non-Fermi liquid phases of the background conduction electrons \cite{Kachru:2010dk}. For a recent account of the application of the AdS/CFT correspondence to quantum critical points of the Hubbard model, see \cite{Sachdev:2010uz}.

The basic ingredient in the top-down construction of holographic dimers in \cite{Kachru:2009xf} are D5-branes embedded into an asymptotically AdS$_5 \times S^5$ black brane background such that four of the six D5-brane coordinates wrap a 4-sphere at a constant latitude on the $S^5$, and the remaining two directions form an effective string within the asymptotically AdS$_5$ factor of bulk. D5-branes describe operators in the anti-symmetric representations of the $SU(N)$ gauge group in $\mathcal{N}=4$ SYM theory \cite{Gomis:2006sb}, at large $N$ and large 't~Hooft coupling. Therefore, the single D5-brane considered in section~II.B of \cite{Kachru:2009xf}, which extends between the brane and the AdS boundary, can be recognized as the gravity dual of the Polyakov loop\footnote{To be precise, this is the Polyakov-Maldacena loop, which differs from the Polyakov loop in the fact that the operator in the loop integral contains also a SYM scalar, see \eqref{saddle:action}.} of such operators on spacial $\mathbb{R}^3$. In the present paper, a field theory calculation of this Polyakov loop is presented.

The Polyakov loop of anti-symmetric representations in $\mathcal{N}=4$ SYM theory on spacial $S^3$ has been calculated holographically in \cite{Hartnoll:2006hr}. On the field theory side, it has been calculated using a unitary matrix model in \cite{Grignani:2009ua, Semenoff:2009zz}, which one obtains after integrating out the SYM fields. This is possible, because the fields on a spacial $S^3$ are gapped due to the conformal coupling to the background metric. 

Circular Wilson loops in $\mathcal{N}=4$ SYM, which are half-BPS configurations, are related to a Hermitian matrix model \cite{Erickson:2000af, Drukker:2000rr, Pestun:2007rz}. For anti-symmetric representations, this model has been solved in 
\cite{Yamaguchi:2006tq, Hartnoll:2006is}, and the results are in perfect agreement with the holographic calculation, based on an AdS$_2$ embedding of the D5-brane in AdS$_5$ \cite{Yamaguchi:2006tq, Hartnoll:2006ib}.
Other BPS Wilson loops in Minkowski space have been considered in \cite{Branding:2009fw}.

For completeness, we mention that the D5-brane in the holographic dimer configuration \cite{Kachru:2009xf} connects two spacially separated impurities (of opposite charges) on the boundary $\mathbb{R}^3$. The configuration in the asymptotically AdS$_5$ part is essentially identical to the confining string in AdS at finite temperature \cite{Rey:1998ik, Rey:1998bq, Imamura:1998hf, Brandhuber:1998bs, Albacete:2008dz}. Spinning D5-branes have been considered in \cite{Armoni:2006ux}.

All of the above D5-brane configurations share the common feature that the latitude of the $S^4$ that wraps the $S^5$ is determined by a quantized azimuth angle $\theta_n$ \cite{Camino:2001at},
\begin{equation}
\label{review:n.theta}
	n = \frac{N}\pi \left( \theta_n -\sin\theta_n\cos\theta_n \right) \qquad \qquad  (0\leq n \leq N)~.
\end{equation}
Here, $N$ is the rank of the $SU(N)$ gauge group corresponding to the number of D3s generating the background, and $n$ is the integer fundamental string charge carried by the D5-brane. The integer $n$ also corresponds to the length of the single-column Young diagram that characterizes the anti-symmetric representation of the operator, or equivalently to the number of fermions living at the impurity site \cite{Gomis:2006sb}. In what follows, we shall consider the limit of large $N$ and use 
\begin{equation}
\label{review:nu.def}
	\nu = \frac{n}{N} = \frac{1}\pi \left( \theta -\sin\theta\cos\theta \right)
\end{equation}
as a continuous parameter, dropping the subscript $n$. This formula has an intriguing similarity with formulas describing the fermion occupation number for certain Green's functions in quantum impurity models, such as the multi-channel $SU(N)$ Kondo model \cite{PhysRevB.58.3794} and the nearly critical Heisenberg spin glass \cite{PhysRevB.63.134406}, which can be taken as an indication that these systems are indeed related. 

Let us review the results of the holographic calculations of the Polyakov loop of anti-symmetric representations. For $\mathcal{N}=4$ SYM on $\mathbb{R}^{3,1}$,\footnote{The single $\mathbb{R}$ is the time direction, which is compactified in the Euclidean finite temperature calculation.} which is what we are mostly interested in, the calculation has been done in \cite{Kachru:2009xf} and involves an asymptotically AdS$_5\times S^5$ black brane background. For $S^3\times \mathbb{R}$, the D5-brane is placed in an asymptotically AdS$_5\times S^5$ black hole \cite{Hartnoll:2006hr}. One can easily generalize the calculation to the case of the hyperbolic space $H^3 \times \mathbb{R}$ by placing a D5-brane in an asymptotically AdS$_5\times S^5$ hyperbolic black brane background.\footnote{These three cases should not be confused with the several BPS Wilson loops considered in \cite{Branding:2009fw}. There, the Wilson loops lie within $S^3$ or $H^3$ which, in turn, are subspaces of the Minkowski space where the SYM theory lives.}
The resulting (renormalized) free energy $F$ of the D5-brane, valid for all three cases, can be written as
\begin{equation}
\label{review:Polyakov}
	\beta F = - \frac{N\sqrt{\lambda}}{3\pi^2 \lad^2} \beta \rh \sin^3 \theta~,
\end{equation}
where $\beta$ is the inverse temperature, $\lambda$ is the `t~Hooft coupling and $\lad$ the AdS radius. The horizon radius $\rh$ is related to $\beta$ by
\begin{equation}
\label{review:beta}
	\beta = \frac{2\pi \rh \lad^2}{2 \rh^2 +\kappa \lad^2}~,
\end{equation}
with $\kappa=0,1,-1$ for the three cases of SYM on spacial $\mathbb{R}^3, S^3$ and $H^3$, respectively. 

In the case of spacial $\mathbb{R}^3$, the temperature dependence on the right hand side of \eqref{review:Polyakov} drops out, so that one obtains  
\begin{equation}
\label{review:Polyakov.R3}
	-\beta F = S = \frac{N\sqrt{\lambda}}{3\pi} \sin^3 \theta~,
\end{equation}
where $S$ is the entropy. This is precisely half of the result of the circular Wilson loop \cite{Yamaguchi:2006tq}, which may appear puzzling at first sight. Whereas the Wilson loop is independent of the radius of the circle contour, the Polyakov loop \eqref{review:Polyakov.R3} is independent of the temperature. It follows that, on spacial $\mathbb{R}^3$, the infinite-radius Wilson loop is not the same as the Polyakov loop at zero temperature, contrary to a naive expectation. From the bulk perspective, this fact is easy to explain by a symmetry argument, which we include in appendix~\ref{Wilson.Polyakov}.

The general result for the entropy is found to be 
\begin{equation}
\label{review:Polyakov.entropy}
	S = \frac{N\sqrt{\lambda}}{3\pi} \sin^3 \theta \left( 1- \frac{\kappa}{2\hat{r}^2} \right)^{-1}~,
\end{equation}
where $\hat{r}=\rh/\lad$. For $\kappa=1$, $\hat{r}$ is restricted to values $\hat{r}\geq 1$ by the instability of the AdS black hole under the Hawking-Page phase transition. It follows that one cannot take the zero temperature limit in this case. The dual SYM theory lives on a spacial $S^3$ of radius $\lad$.
For $\kappa=-1$, one has $\hat{r}\geq 1/\sqrt{2}$ \cite{Cai:2001jc}, with the equal sign giving an extremal hyperbolic black brane. The dual SYM theory lives on a spacial hyperbolic $H^3$ of length scale $\lad$.

In this paper, we shall present a field theoretic calculation of \eqref{review:nu.def} and \eqref{review:Polyakov.R3}. We  exploit the interpretation of the Polyakov loop of an anti-symmetric representation as a quantum impurity model of localized fermions coupling to a background of SYM fields \cite{Gomis:2006sb}. In section~\ref{saddle}, we present the impurity action corresponding to the Polyakov loop and derive the saddle point equations that hold in the large-$N$ limit. In section~\ref{scaling}, we consider a scaling regime, in which the self energy dominates Dyson's equation for the Green's function. The scaling solution is found following closely \cite{Sachdev:2010uj}, but allowing for particle-hole asymmetry in the Green's function. We show that the azimuth angle $\theta$ is related to the spectral asymmetry angle in the fermion Green's function \cite{PhysRevB.58.3794}, and that the impurity solution dual to the D5-brane system is characterized by $\Delta=1/2$ for the fermion scaling dimension, in order for \eqref{review:nu.def} to hold. This value is a limiting case of the saddle point solution, in which the saddle point equations can be solved exactly. We shall do this in section~\ref{exact.sol}, where both \eqref{review:nu.def} and the impurity entropy \eqref{review:Polyakov.R3} will be obtained, the latter up to a numerical factor. Finally, section~\ref{concs} contains the conclusions.

%% file: impurity.tex
\section{Impurity Action and Saddle Point Equations}
\label{saddle}

The single D5-brane at finite temperature is the gravity dual of an impurity operator in the antisymmetric representation, coupled to an $\mathcal{N}=4$ SYM background. The action of this impurity model, in the imaginary time formulation, is given by \cite{Gomis:2006sb}
\begin{align}
\notag
	I &= \int \rmd^3 x \rmd \tau \Lag_{\mathcal{N}=4} + \int \rmd \tau \Lag_{\text{imp}}\\
\label{saddle:action}
	 \Lag_{\text{imp}} &= \chi^\dagger_a \partial_\tau \chi^a + i \chi^\dagger_a \tA^a_b \chi^b 
	 + i\mu (\chi^\dagger_a \chi^a - n)~,
\end{align}
where $\tA$ denotes the combination $\tA(\tau) = A_\tau(0,\tau) + v^I \phi_I(0,\tau)$ of SYM fields in the adjoint of $SU(N)$, $v_I$ is a unit 6-vector. The impurity fermions $\chi$ are in the fundamental representation of $SU(N)$, and their occupation number is fixed to $n$ by the Lagrange multiplier term.

We consider the impurity model \eqref{saddle:action} in the large $N$ limit and try to find a saddle-point solution in a simple scaling form along the lines of \cite{Sachdev:2010uj}. First, we integrate out the SYM fields treating them as a background to which the impurity fields couple. Decompose the fields $\tA$ as
\begin{equation}
\label{saddle:ta.decomp}
	\tA^a_b = (t^c)^a_{\;b} \tA_c~,
\end{equation}
where $t^c$ are the $SU(N)$ adjoint generators, which satisfy 
\begin{equation}
\label{saddle:t.complete}
	(t^c)^a_{\;b} (t^c)^{a'}_{\;b'} = 
	\frac12 \left( \delta^a_{b'} \delta^{a'}_b 
		-\frac1N \delta^a_b \delta^{a'}_{b'}\right)~.
\end{equation}
We are interested in solutions in which the correlations of $\tA$ have the diagonal form\footnote{We write the prefactor motivated by the fact that each gauge field carries a factor of $\gYM$, and $\lambda = N \gYM^2$. This will remove the explicit $N$-dependence from the saddle point equations, but is otherwise irrelevant.} 
\begin{equation}
\label{saddle.D.def}
	\vev{\tA_c(\tau)\tA_{c'}(\tau')} = \frac{\lambda}N \delta_{cc'} D(\tau-\tau')~,
\end{equation}
where $D(\tau)$ is a bosonic (imaginary time) Green's function.
Then, using \eqref{saddle:t.complete} and neglecting the $1/N$ term (or considering a $U(N)$ gauge group), we are left with the following path integral over the impurity fields
\begin{equation}
\label{saddle:Z}
\begin{aligned}
	Z &= \int \mathcal{D}\chi^a \mathcal{D}\mu \, \exp \left\{ 
	-\int \rmd \tau [\chi^\dagger_a (\partial_\tau +i\mu) \chi^a -i\mu n] \right.\\
	&\quad \left.
	+\frac{\lambda}{2N} \int \rmd\tau\rmd\tau' D(\tau-\tau') 
	\chi^\dagger_a(\tau) \chi^\dagger_b(\tau') \chi^b(\tau) \chi^a(\tau') \right\}~.
\end{aligned}
\end{equation}
This partition function is, except for the occupation number, identical to the impurity model considered in \cite{Sachdev:2010uj}. Hence, following \cite{Sachdev:2010uj}, we introduce the Green's function as\footnote{For later convenience, we stick with the standard notation $G(i\omega_n)$ for the Fourier transform of $G(\tau)$ \cite{Bruus2004}.} 
\begin{equation}
\label{saddle:Green}
	G(\tau) \delta^a_b = - \vev{\mathcal{T} \chi^a(\tau)\chi^\dagger_{b'}(0)}~,
	\qquad
	G(i\omega_n) = \int\limits_0^\beta \rmd\tau \e{i\omega_n\tau} G(\tau)~,
\end{equation}
where $\omega_n=(2n+1)\pi/\beta$ are the fermionic Matsubara frequencies.
Then, in the large-$N$ limit, the following set of saddle point equations have to be solved,
\begin{subequations}
\label{saddle:eqs}
\begin{align}
\label{saddle:eqs1}
	G(i\omega_n) &= \frac1{i\omega_n - \chp-\Sigma(i\omega_n)}~, \\
\label{saddle:eqs2}
	\Sigma(\tau) &= \lambda D(\tau) G(\tau)~, \\
\label{saddle:eqs3}
	G(\tau\to 0^-) &= \frac{n}N = \nu~,
\end{align}
\end{subequations}
where $\chp$ is the saddle point value of $i\mu$. 

%%%%%%%%%%%%%%%%%%%%%%%%%%%%%%%%%%%%%%%%%%%%%%%%%%%%%%%%%%%%%%%%%%%%%%%%%%%%%%%%%%%%%%%%%%%%%%%%%%%%%%%%%%%%%%%%%%%%%%%%%%
\section{Solution in the Scaling Regime}
\label{scaling}
\subsection{Scaling Solution}
\label{scaling:sol}
 
We shall now find a simple analytic solution for the system \eqref{saddle:eqs} under the assumption that $D(\tau)$ has a  conformally invariant form. The idea is to generalize the calculation in \cite{Sachdev:2010uj} by assuming a particle-hole asymmetric form of $G(\tau)$. Indeed, we find that the calculation remains nearly unchanged. 
We start with the Green's functions with conformal form
\begin{align}
\label{saddle:Dtau}
	D(\tau) &= A_0 \beta^{-2\Delta_0} \left[\frac{\pi}{\sin(\pi\ttau)} \right]^{2\Delta_0}
	\qquad (0<\tau<\beta)~, \\
\label{saddle:Gtau}
	G(\tau) &=-A\beta^{-2\Delta} \e{\alpha\ttau}\left[\frac{\pi}{\sin(\pi\ttau)}\right]^{2\Delta}
	\qquad (0<\tau<\beta)~,
\end{align}
where we have introduced $\ttau=\tau/\beta$, $\Delta$ and $\Delta_0$ are critical exponents, and $\alpha$ parameterizes the particle-hole asymmetry \cite{PhysRevB.58.3794}. The ans\"atze \eqref{saddle:Dtau} and \eqref{saddle:Gtau} are valid for low energies (in units of some cut-off), or equivalently for large $\lambda$. This is just the supergravity regime. 
The normalization constants $A$ and $A_0$ carry physical dimensions $L^{2\Delta}$ and $L^{2\Delta_0-2}$, respectively, to be consistent with \eqref{saddle:eqs}.\footnote{Notice that $G(\tau)$ and $G(i\omega_n)$, and similarly for $D$ and $\Sigma$, do not have the same dimension, because of the definition of the Fourier transform in \eqref{saddle:Green}.}
The expressions of $D(\tau)$ and $G(\tau)$ for negative $\tau$ follow from the usual periodicity relations for bosons and fermions, respectively, but we do not need them here. The self energy $\Sigma(\tau)$ simply follows from \eqref{saddle:eqs2}.

The Fourier transform of \eqref{saddle:Gtau} can be found with the help of formula 3.631 of \cite{Gradshteyn}, which formally requires $\Delta<1/2$. One obtains
\begin{equation}
\label{saddle:Gom}
	G(i\omega_n) =-iA \frac{(-1)^n \beta^{1-2\Delta} (2\pi)^{2\Delta} \Gamma(1-2\Delta) \e{\alpha/2}}{%
	\Gamma\left(1-\Delta +\frac{\tom_n-i\alpha}{2\pi}\right) 
	\Gamma\left(1-\Delta -\frac{\tom_n-i\alpha}{2\pi}\right) }~,
\end{equation}
where $\tom_n=\beta \omega_n = (2n+1)\pi$. Similarly, Fourier transforming \eqref{saddle:eqs2} yields 
\begin{equation}
\label{saddle:Sigma.om}
	\Sigma_{\text{sing}}(i\omega_n) =-i\lambda A A_0 \frac{(-1)^n \beta^{1-2\Delta-2\Delta_0} (2\pi)^{2\Delta+2\Delta_0} \Gamma(1-2\Delta-2\Delta_0) \e{\alpha/2}}{\Gamma\left(1-\Delta-\Delta_0 +\frac{\tom_n-i\alpha}{2\pi}\right)
\Gamma\left(1-\Delta -\Delta_0-\frac{\tom_n-i\alpha}{2\pi}\right) }~,
\end{equation}
where the subscript indicates that this is only the singular low frequency part. To satisfy Dyson's equation \eqref{saddle:eqs1} in the scaling regime, in which the self energy dominates the denominator of $G(i\omega_n)$, we must impose 
\begin{equation}
\label{saddle:dyson}
	G(i\omega_n)\Sigma_{\text{sing}}(i\omega_n) = -1~.
\end{equation}
This equation can indeed be satisfied for all $n$. Matching the exponent of $\beta$ yields\footnote{From this relation one realizes that \eqref{saddle:Sigma.om} is the singular part of the self energy. The integrals in the Fourier transforms \eqref{saddle:Gom} and \eqref{saddle:Sigma.om} require $\Delta<1/2$ and $\Delta+\Delta_0<1/2$, respectively, which is inconsistent with \eqref{saddle:match.delta}. Therefore, we take $\Delta<1/2$ and drop a cutoff-dependent high frequency piece in $\Sigma$.}
\begin{equation}
\label{saddle:match.delta}
	2\Delta + \Delta_0 =1~.
\end{equation} 
Then, \eqref{saddle:dyson} reduces to a simple relation between the prefactors,
\begin{equation}
\label{saddle:match.A}
	\lambda A^2 A_0 \frac{2\pi}{\Delta_0 \sin(\pi\Delta_0)} \e{\alpha} 
	\left[\cosh\alpha-\cos(\pi\Delta_0)\right] =1~.
\end{equation}
It is a simple check that one obtains equation (25) of \cite{Sachdev:2010uj} in the particle-hole symmetric case, $\alpha=0$. 
Thus, we have shown that the saddle point equations \eqref{saddle:eqs1} and \eqref{saddle:eqs2} have solutions with a conformally invariant structure also in the particle-hole asymmetric case. What remains to be done is to impose \eqref{saddle:eqs3} relating the occupation number $\nu$ to the asymmetry parameter $\alpha$.

%%%%%%%%%%%%%%%%%%%%%%%%%%%%%%%%%%%%%%%%%%%%%%%%%%%%%%%%%%%%%%%%%%%%%%%%%%%%%%%%%%%%%%%%%%%%%%%%%%%%%%%%%%%%%%%%%%%%%%%%%%

\subsection{Retarded Green's Function and Spectral Density}

Before proceeding with \eqref{saddle:eqs3}, let us calculate the retarded Green's function $G^R(\omega)$, and the corresponding spectral density. This is done by analytic continuation $i\omega_n\to \omega$ of $G(i\omega_n)$ in such a way that the function $G(\omega)$ for complex $\omega$ is analytic and has no zeros in the complex upper half plane. We start by rewriting \eqref{saddle:Gom} as 
\begin{equation}
\label{saddle:Gom2}
\begin{aligned}
	G(i\omega_n) &= -iA \frac{(-1)^n \beta^{1-2\Delta} (2\pi)^{2\Delta} \Gamma(1-2\Delta)}{\pi^2} \e{\alpha/2} 
	\Gamma\left(\Delta +i\frac{i\tom_n+\alpha}{2\pi}\right) \\
	&\quad \times
	\Gamma\left(\Delta -i\frac{i\tom_n+\alpha}{2\pi}\right)
	\sin \left(\pi\Delta +i\frac{i\tom_n+\alpha}{2}\right)
	\sin \left(\pi\Delta -i\frac{i\tom_n+\alpha}{2}\right)~.
\end{aligned}
\end{equation}
If we simply replaced $i\tom_n\to\tom$, we would see that the first $\sin$ function, $\sin(\pi\Delta+i\ldots)$, has zeros at $\tom= -\alpha+2\pi i(\Delta+k)$, for any integer $k$. The zeros for $k\geq0$ compensate the poles of the first Gamma function, $\Gamma(\Delta+i\ldots)$, while the remaining zeros have $\im \tom<0$ (remember $\Delta<1/2$), so we keep these functions as they are. The poles of the second Gamma function, $\Gamma(\Delta-i\ldots)$, need not be compensated, as they lie in the lower half plane, but the second $\sin$ function has zeros at $\tom= -\alpha-2\pi i(\Delta+k)$, so that $\im\tom>0$ for $k<0$. Hence, we eliminate these zeros by writing 
\[ \sin \left(\pi\Delta -i\frac{i\tom_n+\alpha}{2}\right) = \sin \left(\pi\Delta +n\pi +\frac{\pi}{2} -i\frac{\alpha}{2}\right) = (-1)^n \cos\left(\pi\Delta -i\frac{\alpha}{2}\right)~. \]
Thus, we obtain, for real $\omega$,
\begin{equation}
\label{saddle:GR}
\begin{aligned}
	G^R(\omega) &= -iA \frac{\beta^{1-2\Delta} (2\pi)^{2\Delta} \Gamma(1-2\Delta)}{\pi^2} \e{\alpha/2}  
	\left|\Gamma\left(\Delta +i\frac{\tom+\alpha}{2\pi}\right)\right|^2 \\
	&\quad \times
	\sin \left(\pi\Delta +i\frac{\tom+\alpha}{2}\right)
	\cos \left(\pi\Delta -i\frac{\alpha}{2}\right)~.
\end{aligned}
\end{equation}
Let us elaborate this expression a bit further. First, to recover the $\omega\to 0$ singularity in the zero temperature limit, one considers $|\omega|\gg 1/\beta$, \ie $|\tom|\gg 1$. Hence, with the help of formula 8.328 of \cite{Gradshteyn} one obtains\footnote{Notice that one has to consider separately the cases $\tom>0$ and $\tom<0$, but both give rise to the same result.}
\begin{equation}
\label{saddle:GR.asy}
	G^R(\omega) \sim 2A\,\Gamma(1-2\Delta) \e{\alpha/2}  \cos\left(\pi\Delta -i\frac{\alpha}{2}\right)
	\frac{\e{-i\pi \Delta}}{\omega^{1-2\Delta}}~.
\end{equation}
Let us rewrite this as 
\begin{equation}
\label{saddle:GR.asy1}
	G^R(\omega) \sim h(\vt) \frac{\e{-i\pi \Delta-i\vt}}{\omega^{1-2\Delta}}~,
\end{equation}
where the angle $\vt$ parameterizes the spectral asymmetry \cite{PhysRevB.58.3794} and has been introduced by 
\begin{equation}
\label{saddle:vt.def}
	\cos\left(\pi\Delta -i\frac{\alpha}{2}\right) 
	= \left| \cos\left(\pi\Delta -i\frac{\alpha}{2}\right) \right| \e{-i\vt}~.
\end{equation}
Positivity of the spectral function (see below) requires $|\vt|\leq \pi\Delta$, and \eqref{saddle:vt.def} also yields the useful relation
\begin{equation}
\label{saddle:alpha.vt}
	\e{\alpha} = \frac{\sin(\pi\Delta-\vt)}{\sin(\pi\Delta+\vt)}~.
\end{equation}
The real and positive prefactor $h(\vt)$ in \eqref{saddle:GR.asy1} is 
\begin{equation}
\label{saddle:h.vt}
	h(\vt) = 2A\, \Gamma(1-2\Delta) \e{\alpha/2} \left| \cos\left(\pi\Delta -i\frac{\alpha}{2}\right) \right|~.
\end{equation}
The scaling behaviour \eqref{saddle:GR.asy1} implies that the spectral density behaves as  
\begin{equation}
\label{saddle:spec.fun}
	\rho(\omega) = -2 \im G^R(\omega) \sim
	\begin{cases} 
		\frac{C_+}{\omega^{1-2\Delta}} \qquad\qquad  &\omega>0~,\\
		\frac{C_-}{(-\omega)^{1-2\Delta}} &\omega<0~,
	\end{cases}
\end{equation}
where 
\begin{equation}
\label{saddle:C.pm}
	C_{\pm} = 2h(\vt) \sin(\pi\Delta\pm \vt)~.
\end{equation}

Let us conclude by performing a similar calculation for the bosonic background Green's function $D(\tau)$. 
Fourier transforming \eqref{saddle:Dtau} gives
\begin{equation}
\label{saddle:Dom}
\begin{aligned}
	D(i\nu_n) &= A_0 \frac{(-1)^n \beta^{1-2\Delta_0} (2\pi)^{2\Delta_0} \Gamma(1-2\Delta_0)}{\pi^2} 
	\Gamma\left(\Delta_0 -\frac{\tnu_n}{2\pi}\right) \Gamma\left(\Delta_0 +\frac{\tnu_n}{2\pi}\right)\\
	&\quad \times
	\sin \left(\pi\Delta_0 -\frac{\tnu_n}{2}\right)
	\sin \left(\pi\Delta_0 +\frac{\tnu_n}{2}\right)~,
\end{aligned}
\end{equation}
where $\nu_n=2n\pi/\beta$ are the bosonic Matsubara frequencies, and $\tnu_n=\beta\nu_n$. 
The analytic continuation to complex frequencies, $i\nu_n\to \omega$, yields for real $\omega$
\begin{equation}
\label{saddle:DR}
	D^R(\omega) = A_0 \frac{\beta^{1-2\Delta_0} (2\pi)^{2\Delta_0} \Gamma(1-2\Delta_0) \sin(\pi\Delta_0)}{\pi^2}
	\left|\Gamma\left(\Delta_0 +i\frac{\tom}{2\pi}\right) \right|^2
	\sin \left(\pi\Delta_0 +i\frac{\tom}{2}\right)~.
\end{equation}
Thus, the spectral density is found to be
\begin{equation}
\label{saddle:spec.D}
	\rho_D(\omega) = -2\im D^R(\omega) = - A_0 \frac{\beta^{1-2\Delta_0} (2\pi)^{2\Delta_0}}{\pi \Gamma(2\Delta_0)} 	\left|\Gamma\left(\Delta_0 +i\frac{\tom}{2\pi}\right) \right|^2 \sinh\left(\frac{\tom}{2}\right)~, 
\end{equation}
and its low-temperature scaling behaviour is recovered as
\begin{equation}
\label{saddle:spec.D.asy}
	\rho_D(\omega) \sim - \sgn(\omega) \frac{C_0}{|\omega|^{1-2\Delta_0}}~,\qquad 
	C_0 = A_0 \frac{2\pi}{\Gamma(2\Delta_0)}~.
\end{equation}

%%%%%%%%%%%%%%%%%%%%%%%%%%%%%%%%%%%%%%%%%%%%%%%%%%%%%%%%%%%%%%%%%%%%%%%%%%%%%%%%%%%%%%%%%%%%%%%%%%%%%%%%%%%%%%%%%%%%%%%%%%

\subsection{Fermion Occupation Number}

Let us now return to \eqref{saddle:eqs3}. Imposing \eqref{saddle:eqs3} is not straightforward, because the scaling forms of the solution do not hold for small $\tau$, \ie in the high energy regime. The clearest indication of this is the fact that the standard relation $G(0^+)+ G(\beta^-) =-1$ cannot be satisfied by \eqref{saddle:Gtau}. 
Therefore, we shall follow a more indirect route \cite{PhysRevB.58.3794, PhysRevB.63.134406, Abrikosov}. 

Let us consider the low temperature limit. At zero temperature, the occupation number can be written as 
\begin{equation} 
\label{saddle:nu}
	\nu = i G^F(t\to 0^-) = \frac{i}{2\pi} \int\limits_{-\infty}^\infty \rmd \omega\, G^F(\omega) \e{i\omega 0^+}~,
\end{equation}
where $G^F$ is the zero temperature Feynman Green's function. Using the analytic continuation of \eqref{saddle:eqs1}, one can rewrite \eqref{saddle:nu} as 
\begin{equation} 
\label{saddle:nu1}
	\nu = \frac{i}{2\pi} \Pval \int\limits_{-\infty}^\infty \rmd \omega\, \e{i\omega 0^+} 
		\partial_\omega \ln G^F(\omega) 
	- \frac{i}{2\pi} \Pval \int\limits_{-\infty}^\infty \rmd \omega\, 
	\e{i\omega 0^+} G^F(\omega) \partial_\omega \Sigma^F(\omega)~.
\end{equation}
Here, $\Pval$ denotes the principal value of the integral, defined as
\[ \Pval \int\limits_{-\infty}^\infty \rmd \omega = \lim\limits_{\eta\to0^+} 
	\left( \int\limits_{-\infty}^{-\eta} \rmd \omega + \int\limits_{\eta}^{\infty} \rmd \omega \right)~.
\]
The first integral in \eqref{saddle:nu1} is done by using the relation between $G^F$ and $G^R$ and exploiting the analyticity and scaling properties \eqref{saddle:GR.asy} of the latter, 
\[
	\Pval \int\limits_{-\infty}^\infty \rmd \omega\, \e{i\omega 0^+} \partial_\omega \ln G^F(\omega) = 
	\Pval \int\limits_{-\infty}^\infty \rmd \omega\, \e{i\omega 0^+} \partial_\omega \ln G^R(\omega) 
	- \int\limits_{-\infty}^{0^-} \rmd \omega\, \e{i\omega 0^+} \partial_\omega \ln 
		\frac{G^R(\omega)}{G^R{}^\ast(\omega)}~.
\]
The first integral on the right hand side is done by closing the integration contour with a tiny half circle above the $\omega=0$ pole of \eqref{saddle:GR.asy1} and another half circle at infinity in the upper half plane. Using the analyticity of the integrand in the upper half plane, the nonzero value stems from the half circle around the pole. The second integral on the right hand side is straightforward. Using $\arg G^R(0^-)= \pi\Delta-\pi-\vt$, which one finds from \eqref{saddle:GR.asy1}, and the general property $\arg G^R(-\infty)= -\pi$, one obtains
\begin{align}
\notag
	\Pval \int\limits_{-\infty}^\infty \rmd \omega\, \e{i\omega 0^+} \partial_\omega \ln G^F(\omega) 
	&= \frac12 (2\pi i) \res_{\omega=0} \left(\frac{2\Delta-1}{\omega}\right) 
		-2i \left[ \arg G^R(0^-) - \arg G^R(-\infty) \right] \\
\label{saddle:nu2}
	&= \pi i (2 \Delta -1) -2i (\pi\Delta-\vt) = -i (\pi -2\vt)~.
\end{align}

The second integral in \eqref{saddle:nu1} is the difficult part, but we shall use a trick \cite{PhysRevB.63.134406} to find the result without explicitly doing the integral. From the existence of the Luttinger-Ward functional 
\begin{equation}
\label{saddle:LW}
	\LW = \frac{\lambda}2 \int \rmd \tau\, D(\tau) G(\tau) G(-\tau)~,  
\end{equation} 
such that \eqref{saddle:eqs2} follows from $\Sigma(\omega) = \delta\LW/\delta G(\omega)$, one might naively conclude that the second integral in \eqref{saddle:nu1} vanishes \cite{Abrikosov}. However, this conclusion is spoilt by an anomaly, because there does not exist a regularization such that the formal invariance of $\LW$ under frequency shifts can be used. The reason lies in the definition of the principle value of the frequency integral. From this we can deduce that the integral should depend on the coefficients $C_0$ and $C_\pm$ of the spectral functions in the positive and negative frequency regions, eqs.~\eqref{saddle:spec.D.asy} and \eqref{saddle:spec.fun}, in a form that is dictated by the form of $\LW$. Hence, we write
\begin{equation} 
\label{saddle:anom.int}
	\frac{i}{2\pi} \Pval \int\limits_{-\infty}^\infty \rmd \omega\, 
	\e{i\omega 0^+} G^F(\omega) \partial_\omega \Sigma^F(\omega) = B\, \lambda C_0 (C_+^2 -C_-^2)
\end{equation}
with some coefficient $B$ that is to be determined by imposing the boundary condition $\nu=0$ for $\vt=\pi\Delta$ (or $\nu=1$ for $\vt=-\pi\Delta$). 

From \eqref{saddle:spec.D.asy}, \eqref{saddle:C.pm} and \eqref{saddle:h.vt}, and using the relations \eqref{saddle:match.delta} and \eqref{saddle:match.A}, one finds
\begin{equation}
\label{saddle:C.comb}
	C_0 (C_+^2 -C_-^2) = \frac{8\Delta_0 [\Gamma(\Delta_0)]^2 \sin^2(\pi\Delta_0)}{\lambda \Gamma(2\Delta_0)} 
	\sin(2\vt)~.
\end{equation}
Then, substituting the results \eqref{saddle:nu2}, \eqref{saddle:anom.int} and \eqref{saddle:C.comb} into \eqref{saddle:nu1} and imposing the boundary conditions at $\vt=\pm \pi \Delta$, one determines
\begin{equation}
\label{saddle:B}
	B = \frac{\Gamma(2\Delta_0)}{16[\Gamma(\Delta_0)]^2 \sin^3(\pi\Delta_0)}~.
\end{equation}
Finally, one obtains
\begin{equation}
\label{saddle:nu.final}
	\nu = \frac12 -\frac{\vt}{\pi} -\frac{\Delta_0}{2 \sin(\pi\Delta_0)} \sin(2\vt)~.
\end{equation}
This formula, together with the relation \eqref{saddle:alpha.vt}, completes the solution of the saddle point equations \eqref{saddle:eqs}.

Let us compare our result with the D5-brane formula \eqref{review:nu.def}. Noting that $0\leq\theta\leq\pi$, while $-\pi\Delta\leq\vt\leq \pi\Delta$, we propose that the D5-brane system corresponds to the limiting case 
\begin{equation}
\label{saddle:D5.rel}
	\Delta=1/2~, \qquad \Delta_0=0~, \qquad \vt = \pi/2-\theta~.
\end{equation}
Hence, the azimuth angle $\theta$ is related to the spectral asymmetry angle $\vt$. We note that the crucial relations of the solution remain valid in the $\Delta_0\to0$ limit, if we express the parameter $\alpha$ in terms of the spectral asymmetry angle $\vt$. Specifically, using \eqref{saddle:alpha.vt} and \eqref{saddle:match.delta}, the matching condition \eqref{saddle:match.A} can be rewritten as
\begin{equation}
\label{saddle:match.A2}
	\frac{2\pi^2\lambda A^2 A_0\e{\alpha} }{\cos(2\vt)+\cos(\pi\Delta_0)} \frac{\sin(\pi\Delta_0)}{\pi\Delta_0} =1~.
\end{equation}
Remarkably, taking the $\Delta_0\to0$ limit in \eqref{saddle:nu.final}, we recover precisely \eqref{review:nu.def}. Hence, we conclude that our scaling solution with the parameters \eqref{saddle:D5.rel} describes the D5-brane impurity.

As a final remark, we note that the choice of a spacial $\mathbb{R}^3$ has entered very little in the calculation so far, due to the fact that the impurity is point-like. The only place where it may have entered is in the validity of the conformal ansatz for the Green's functions \eqref{saddle:Dtau} and \eqref{saddle:Gtau}. The fact that \eqref{review:nu.def} is independent of the choice of the space suggests that the ansatz is indeed valid in the supergravity regime (large $N$ and $\lambda$).  

%%%%%%%%%%%%%%%%%%%%%%%%%%%%%%%%%%%%%%%%%%%%%%%%%%%%%%%%%%%%%%%%%%%%%%%%%%%%%%%%%%%%%%%%%%%%%%%%%%%%%%%%%%%%%%%%%%%%%%%%%%

\section{Exact Solution for the Limiting Case $\Delta_0=0$}
\label{exact.sol}
\subsection{Solution of the Saddle Point Equations}

Let us solve the saddle point equations \eqref{saddle:eqs} exactly for the case $\Delta_0=0$. In this regime, the bosonic Green's function $D(\tau)$ takes the simple form 
\begin{equation}
\label{saddle:D.special}
	D(\tau) = \frac{c^2}{\beta^2}.
\end{equation}
where $c$ is some positive dimensionless constant in the case of SYM on spacial $\mathbb{R}^3$. Since we are considering a background CFT, there is no other length scale except the temperature, so that the $1/\beta^2$ factor follows simply from dimensionality. One might worry that making $A$ and $A_0$ in \eqref{saddle:Dtau} and \eqref{saddle:Gtau} temperature dependent would spoil the derivation of the scaling solution, but this is not the case. In fact, $A$ and $A_0$ are entirely dummy symbols, related to each other by \eqref{saddle:match.A}, and the relation \eqref{saddle:match.delta} remains necessary in order to satisfy \eqref{saddle:dyson} for all values of $\omega_n$. We note that one can formally include also the cases of spacial $S^3$ and $H^3$ (of radius $\lad$) by considering $c$ to be a function of the dimensionless $\beta/\lad$. The calculation in this subsection would not be affected by this.

Hence, after analytic continuation, \eqref{saddle:eqs1} and \eqref{saddle:eqs2} reduce to
\begin{equation}
\label{saddle:G.eq}
	G(\omega) =\frac{1}{\omega -\chp -\hlambda G(\omega)}~,
\end{equation}
where we have defined 
\begin{equation}
\label{saddle:hlambda}
	\hlambda = c^2 \frac{\lambda}{\beta^2}~.
\end{equation}
Equation~\eqref{saddle:G.eq} is easily solved, and one finds the retarded Green's function 
\begin{equation}
\label{saddle:G.sol}
	G(\omega) =\frac{1}{2\hlambda}\left\{ \omega -\chp - \left[(\omega-\chp)^2 -4 \hlambda\right]^{1/2} \right\}~.
\end{equation}
The spectral density $\rho(\omega) = -2 \im G(\omega)$ is obtained as 
\begin{equation}
\label{saddle:rho}
	\rho(\omega) = \begin{cases} 
	\frac2{\sqrt{\hlambda}} \sqrt{1-\frac{(\omega-\chp)^2}{4 \hlambda}} \qquad &
	\text{for $|\omega -\chp|< 2\sqrt{\hlambda}$,}\\
	0 & \text{otherwise.} \end{cases}
\end{equation}
It is easily checked that \eqref{saddle:rho} satisfies the standard normalization condition
\begin{equation}
\label{saddle:rho.norm}
	\int\limits_{-\infty}^\infty \frac{\rmd \omega}{2\pi} \rho(\omega) = 1~.
\end{equation}

Consider now the occupation number constraint \eqref{saddle:eqs3}. We start from the following representation of the imaginary-time Green's function \cite{Bruus2004}
\begin{equation}
\label{saddle:G.rho}
	G(\tau) = - \int\limits_{-\infty}^\infty \frac{\rmd \varepsilon}{2\pi} \rho(\varepsilon) 
	\frac{\e{-\tau\varepsilon}}{1+\e{-\beta\varepsilon}}~.
\end{equation}
Using the relations $G(0^-)=-G(\beta^-)$, and $G(0^+) + G(\beta^-)=-1$, we rewrite \eqref{saddle:eqs3} as
\begin{equation}
\label{saddle:nu.eq1}
	\nu -\frac12 = \frac12 \left[ G(0^+) - G(\beta^-) \right]~.
\end{equation}
Then, substituting \eqref{saddle:G.rho}, changing the integration variable by $\varepsilon = \chp +2 \sqrt{\hlambda}\, x$ and defining 
\begin{equation}
\label{saddle:muhat}
	\hmu = \frac{\chp}{2 \sqrt{\hlambda}}~,
\end{equation}
we end up with the following integral,
\begin{align}
\notag 
	\nu-\frac12 &= - \frac{1}{\pi} \int\limits_{-1}^1\rmd x \sqrt{1-x^2} \tanh[\beta \sqrt{\hlambda}(\hmu+x)]\\
\label{saddle:nu.eq2}
	&= -\frac{2}{\pi} \int\limits_0^1\rmd x \sqrt{1-x^2} 
	\frac{\tanh(\beta \sqrt{\hlambda} \hmu)}{1+ \frac{\sinh^2(\beta \sqrt{\hlambda} x)}{\cosh^2(\beta \sqrt{\hlambda}\hmu)}}~.
\end{align}
In the limit of large 't~Hooft coupling, which is what counts for the comparison with the D5-brane system, this integral can be easily done. Keeping the temperature fixed and considering large $\lambda$, one has  
\[ 	\lim_{\lambda\to\infty} \frac{\sinh^2(\beta \sqrt{\hlambda} x)}{\cosh^2(\beta \sqrt{\hlambda}\hmu)} = 
	\begin{cases} \infty \qquad &\text{for $x>|\hmu|$}\\ 0 &\text{for $x<|\hmu|$} \end{cases}~,
	\qquad 
	\lim_{\lambda\to\infty} \tanh(\beta \sqrt{\hlambda}\hmu) = \sgn \hmu~.
\] 
Therefore, if $|\hmu|\leq 1$, \eqref{saddle:nu.eq2} becomes 
\begin{equation}
\label{saddle:nu.eq3}
	\nu-\frac12 = -\frac{2}{\pi} \sgn\hmu \int\limits_0^{|\hmu|} \rmd x \sqrt{1-x^2}~,
\end{equation}
and the result is 
\begin{equation}
\label{saddle:nu.eq4}
	\nu-\frac12 = 
	\begin{cases} 
	-\frac1{\pi} \left( \arcsin \hmu +\hmu \sqrt{1-\hmu^2} \right)\qquad &\text{for $|\hmu|\leq 1$}~,\\
	-\frac12 \sgn \hmu &\text{for $|\hmu|>1$}~.
	\end{cases}
\end{equation}
Notice that this is continuous at the points $\hmu=\pm 1$.

Remarkably, if we write, for $|\hmu|\leq 1$,
\begin{equation}
\label{saddle:mu.theta.rel}
	\hmu = \sin\vt = \cos\theta~,
\end{equation}
we recover \eqref{saddle:nu.final} with $\Delta_0=0$, which agrees with the D5-brane relation \eqref{review:nu.def}.
Let us emphasize that this result is independent of the space we are considering ($\mathbb{R}^3$, $S^3$ or $H^3$). All we need is a non-zero $c$ (which may place restrictions on the temperature in the cases of spacial $S^3$ and $H^3$) and large $\lambda$.

%%%%%%%%%%%%%%%%%%%%%%%%%%%%%%%%%%%%%%%%%%%%%%%%%%%%%%%%%%%%%%%%%%%%%%%%%%%%%%%%%%%%%%%%%%%%%%%%%%%%%%%%%%%%%%%%%%%%%%%%%%

\subsection{Impurity Entropy}
We shall now derive the impurity entropy for the case of spacial $\mathbb{R}^3$ and show that, up to the fact that we have the undetermined constant $c$ from \eqref{saddle:D.special}, the result is precisely \eqref{review:Polyakov.R3}. Reversely, we can use \eqref{review:Polyakov.R3} to deduce that $c=1/4$.  

To calculate the impurity entropy, we make use of the thermodynamic identity 
\begin{equation}
\label{saddle:tdyn.ident}
	\frac{\partial S}{\partial \nu} = - N \frac{\partial \chp}{\partial T}~,
\end{equation}
where $T=1/\beta$ is the temperature. This identity can derived as follows. Differentiate the functional integral \eqref{saddle:Z} with respect to $\nu$ to obtain 
\begin{equation}
\label{saddle:dZ.dnu}
	\frac{\partial Z}{\partial \nu} = Z \vev{N \int\limits_0^\beta \rmd \tau\, i\mu}  = Z N \beta \chp~.
\end{equation}
Then, identifying $Z=\e{-\beta F}$ yields 
\begin{equation}
\label{saddle:dF.dnu}
	\frac{\partial F}{\partial \nu} = - N \chp~,
\end{equation}
and differentiating with respect to the temperature leads to \eqref{saddle:tdyn.ident}. 

From \eqref{saddle:tdyn.ident}, relating $\nu$ to $\hmu$ by \eqref{saddle:nu.eq4} (for $|\hmu|\leq1$) and using \eqref{saddle:muhat} and \eqref{saddle:hlambda} (for fixed $\lambda$), one obtains
\begin{equation}
\label{saddle:dS.dmu}
	\frac{\partial S}{\partial \hmu} = -N \frac{\partial \chp}{\partial T} \frac{\partial \nu}{\partial \hmu} 
	= -N \left[\frac{\partial}{\partial T}\left( 2c \sqrt{\lambda} \hmu T \right) \right]
	\left( -\frac2{\pi} \sqrt{1-\hmu^2} \right) 
	= c \frac{4}{\pi} N \sqrt{\lambda} \hmu \sqrt{1-\hmu^2}~.
\end{equation}
It should be appreciated that the factor $1/\beta^2$ in \eqref{saddle:D.special}, which followed from purely dimensional considerations, is crucial for obtaining this result.

Finally, integrating \eqref{saddle:dS.dmu} with respect to $\hmu$ yields
\begin{equation}
\label{saddle:S.final}
	S = c\frac{4N \sqrt{\lambda}}{3\pi}  \left( 1-\hmu^2\right)^{3/2}~.
\end{equation}
Remarkably, after using \eqref{saddle:mu.theta.rel}, this reproduces \eqref{review:Polyakov.R3} with the value $c=1/4$.

%% file: concs.tex
\section{Conclusions}
\label{concs}

In this paper, we have presented a field theoretic calculation, which reproduces the two main predictions from the D5-brane system dual to the Polyakov loop of anti-symmetric representations in $\mathcal{N}=4$ SYM theory on spacial $\mathbb{R}^3$, for large $N$ and large 't~Hooft coupling, namely the relation \eqref{review:nu.def} between the fermion occupation number and an angular parameter, and the impurity entropy \eqref{review:Polyakov.R3}. It has been shown that the solution relevant for the Polyakov loop has fermion scaling dimension $\Delta=1/2$, and that the azimuth angle of the D5-brane configuration is related to the spectral asymmetry angle in the fermion Green's function in the scaling regime. Combining the value $\Delta=1/2$ ($\Delta_0=0$) with purely dimensional arguments, the form of the background Green's function $D(\tau)$ was deduced. After solving the saddle point equations exactly, the impurity entropy \eqref{review:Polyakov} was obtained up to an unknown numerical factor, but exhibiting specifically the $\sqrt{\lambda}$ enhancement. The comparison of numerical factors shows that the correlator of the SYM field $\tA_a$ is, in the regime considered,
\begin{equation}
\label{concs.D}
	\vev{\tA_a(\tau)\tA_{a'}(\tau')} = \frac{\lambda}{16N\beta^2} \delta_{aa'}~.
\end{equation}
It would be interesting to have an independent confirmation of this result. 

In the case of spacial $S^3$ or $H^3$, which are charachterized by the length scale $\lad$, the constant $c$ in \eqref{saddle:D.special} should be considered as a function of the dimensionless factor $\lad T$. The derivation of the fermion occupation number remains valid in this case, which corresponds nicely to what one finds from the D5-brane, namely that \eqref{review:nu.def} is independent of the choice of the space. What would be changed is the calculation of the impurity entropy, specifically \eqref{saddle:dS.dmu}. One can, of course, use the D5-brane result \eqref{review:Polyakov.entropy} to detemine $c(\lad T)$.

Due to the generality of the saddle point equations that we solved, our calculation should generalize to similar quantum impurity models, such as those reviewed in \cite{Sachdev:2010uj}. Moreover, it may provide a step towards a field theory treatment of two impurities coupled to $\mathcal{N}=4$ SYM theory on spacial $\mathbb{R}^3$, which correspond to the holographic dimer system \cite{Kachru:2009xf}. On the bulk side, the holographic dimers are limited to cases, where the impurities carry opposite ``spin''. It would be very interesting to find the D5-brane configurations that are dual, \eg\ to a lattice of impurities with arbitrary ``spin''.

%% file: wilson_polyakov.tex
\section{The Circular Wilson Loop is Twice the Polyakov Loop}
\label{Wilson.Polyakov}

As stated in the introduction, the result for the circular Wilson loop is precisely twice the result \eqref{review:Polyakov.R3} of the Polyakov loop on spacial $\mathbb{R}^3$. Moreover, whereas the Wilson loop is independent of the circle radius $R$, the Polyakov loop is independent of the temperature $1/\beta$. 
This is somewhat puzzling, because one may consider the limits $R\to \infty$ and $\beta\to \infty$, respectively, both giving rise, naively, to an infinite Wilson line at zero temperature. To understand why the Polyakov loop and the Wilson line at zero temperature are actually not the same, let us consider the embedding of the effective string world-sheet in AdS$_5$ and use the AdS symmetries to relate the Wilson and Polyakov loop configurations to each other. Our arguments are in favour of the regularization that leads to (3.17) in \cite{Branding:2009fw}.

\begin{figure}[t]
\begin{center}
a)\includegraphics[width=0.4\textwidth]{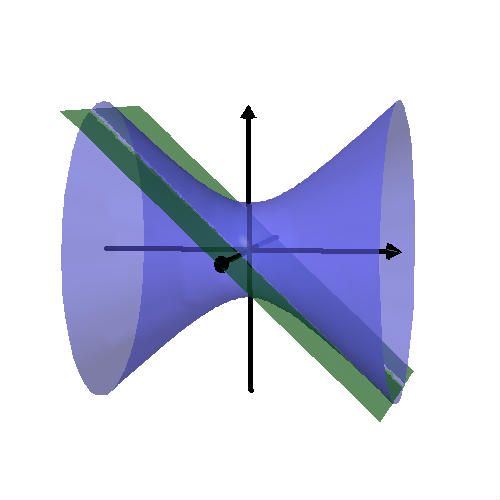}
b)\includegraphics[width=0.4\textwidth]{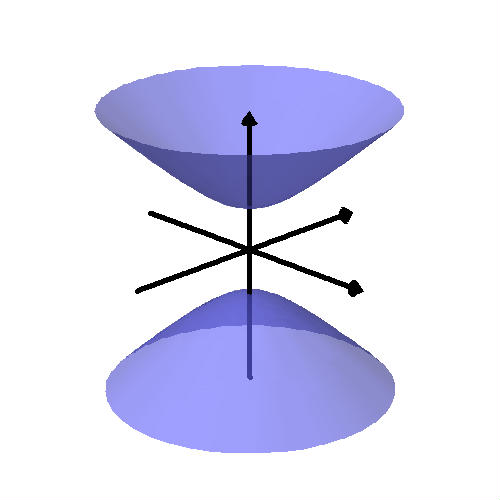}
\end{center}
\caption{\label{AdS.fig}a) A cartoon picture of AdS$_5$ as an embedding in $\mathbb{R}^{4,2}$. The directions $y_1$ (axis to the right), $y_0$ and $y_5$ (axes to the front and top) are shown, and $y_2=y_3=y_4=0$. The intersection of the plane with the hyperboloid is the locus of the horizon. The coordinate system \eqref{review:ads.P.metric} covers only one of the regions of the hyperboloid above or below the plane.\newline
b) Euclidean AdS$_5$ as an embedding in $\mathbb{R}^{5,1}$. The directions $y_0$ and $y_1$ (horizontal axes), and $y_5$ (vertical axis) are shown, and $y_2=y_3=y_4=0$. The horizontal circles should be thought of as $S^4$s. The two branches are disconnected and must be regarded as two copies of EAdS. A Poincar\'e sliced coordinate system similar to \eqref{review:ads.P.metric} (with the Euclidean instead of the Minkowski metric) covers entirely one of the two branches.}
\end{figure}

AdS$_5$ can be considered as a hypersurface in $\mathbb{R}^{4,2}$, defined by 
\begin{equation}
\label{review:ads.embed}
	\eta_{AB} y^A y^B = - y_0^2 + y_1^2 +y_2^2 +y_3^2 +y_4^2 -y_5^2 = -\lad^2~,
\end{equation}
where $\eta_{AB}$ ($A,B=0,\ldots,5)$ is the metric on $\mathbb{R}^{4,2}$, and $\lad$ is the AdS ``radius''. 
Eq.~\eqref{review:ads.embed} defines AdS$_5$ with Lorentzian signature, which is crucial for what follows. 
Poincar\'e-sliced coordinates on AdS$_5$ can be defined by 
\begin{equation}
\label{review:ads.P.coords}
	z = \frac{\lad^2}{y_4+y_5}~,\qquad x^\mu = \frac{z}{\lad} y_\mu\qquad (\mu=0,1,2,3)~,
\end{equation}
giving rise to the induced metric
\begin{equation}
\label{review:ads.P.metric}
	\rmd s^2 = \frac{\lad^2}{z^2} \left(\rmd z^2 + \eta_{\mu\nu} \rmd x^\mu \rmd x^\nu \right)~,
\end{equation}
where $\eta_{\mu\nu}=\diag(-1, 1, 1, 1)$. The definition \eqref{review:ads.P.coords} divides AdS into two coordinate patches, with $z>0$ and $z<0$, respectively, each covering exactly half of AdS$_5$. The boundary is located at $z=0$, and there is a horizon at $|z|=\infty$, beyond which the other coordinate patch should be used. In what follows, we shall work in the patch $z>0$. AdS$_5$ is illustrated in Figure~\ref{AdS.fig}. A global coordinate system exists also, but we will not need it explicitly. 

The main idea is to start with a circular Wilson loop configuration and use the AdS symmetries, which are just rotations and boosts in $\mathbb{R}^{4,2}$, in such a way that half of the effective string world-sheet will be hidden behind the horizon for a Poincar\'e observer.\footnote{The AdS symmetries act as conformal symmetries on the boundary.} Because there exists a global coordinate system in AdS, applying the AdS symmetries does not affect the result for the Wilson loop in global coordinates. However, when switching to Poincar\'e coordinates, only the portion of the world-sheet visible by the observer contributes. For the configuration that corresponds to the Polyakov loop at zero temperature, this is half of the world-sheet. Consequently, the Polyakov loop is precisely half of the Wilson loop. 

Without loss of generality, we can start with a circular Wilson loop of radius $\lad$. Its effective string world-sheet in AdS$_5$ can be written as \cite{Yamaguchi:2006tq}
\begin{equation}
\label{review:wils.l}
	x^1 = \rho \cos \tau~, \quad x^2 = \rho \sin \tau~,\quad x^0=x^3=0~,\quad 
	z = \sqrt{\lad^2-\rho^2}~,\quad 
	(0\leq \rho\leq \lad~, 0\leq \tau <2\pi)~.
\end{equation}
Using \eqref{review:ads.P.coords}, we can find the corresponding $\mathbb{R}^{4,2}$ coordinates,
\begin{equation}
\label{review:wils.l.y}
	y_1 = \frac{\rho \lad\cos \tau}{\sqrt{\lad^2-\rho^2}}~, \quad 
	y_2 = \frac{\rho \lad\sin \tau}{\sqrt{\lad^2-\rho^2}}~, \quad
	y_5 = \frac{\lad^2}{\sqrt{\lad^2-\rho^2}}~, \quad
	y_0=y_3=y_4=0~.
\end{equation}
This is easily recognized as an EAdS$_2$.

\begin{table}[t]
\renewcommand{\arraystretch}{1.5}
\begin{tabular}{|c|m{0.1\textwidth}|c|c|c|c|m{0.19\textwidth}|}
\hline
 \# & Config. & $x_0$    & $x_1$     & $x_2$     & $z$    & Boundary Curve \\ \hline
1   & \multicolumn{1}{l|}{---}              & 	$0$     & $\rho\cos\tau$ & $\rho\sin\tau$ & $\sqrt{\lad^2-\rho^2}$ & 
	circular Wilson loop, $x_1^2 +x_2^2=\lad^2$ 
\\ \hline 
2   & \#1 + \linebreak $(2,4)$ rot & $0$  & $\frac{\rho\cos\tau}{1+\frac{\rho}{\lad}\sin\tau\sin\alpha}$ &
	$\frac{\rho\sin\tau\cos\alpha}{1+\frac{\rho}{\lad}\sin\tau\sin\alpha}$ & 
	$\frac{\sqrt{\lad^2-\rho^2}}{1+\frac{\rho}{\lad}\sin\tau\sin\alpha}$ & 
	circle with centre $(0,-l\tan\alpha)$, radius $\frac{\lad}{\cos\alpha}$ 
\\ \hline 
3   & \#2, \linebreak $\alpha=\pi/2$ & $0$  & $\frac{\rho\cos\tau}{1+\frac{\rho}{\lad}\sin\tau}$ &
	$0$ & $\frac{\sqrt{\lad^2-\rho^2}}{1+\frac{\rho}{\lad}\sin\tau}$ & 
	straight line along $x_1$
\\ \hline 
4   & \#3 +\linebreak $(0,5)$ rot \linebreak $(0,4)$ b. $\alpha=\pi/4$, $\gamma=-1$ & 
	$\frac{\lad(\rho\sin\tau-\lad)}{\sqrt{2}\rho\sin\tau}$ & 
	$\frac{\lad \cot\tau}{\sqrt{2}}$ & $0$ &
	$\frac{\lad \sqrt{\lad^2-\rho^2}}{\sqrt{2}\rho\sin\tau}$ & 
	hyperbola, \newline $\pi<\tau<2\pi$ part of worldsheet is behind the horizon
\\ \hline 	
5   & \#4 +\linebreak $(4,5)$ b. & \multicolumn{4}{m{0.4\textwidth}|}{values of \#4 multiplied by $\sqrt{\gamma^2+1}+\gamma$} &
	limit $\gamma\to \infty$ gives straight line along $x_1$, invisible branch moved to infinity 
\\ \hline
\end{tabular}
\caption{\label{review:tab.trans}Transformation of the circular Wilson loop into a Polyakov loop at zero temperature. We start with the circular Wilson loop as configuration $\# 1$ and apply a series of rotations and boosts, as indicated in the second column for each configuration. Columns 3--6 list the relevant Poincar\'e coordinates for each configuration, and the last column describes the geometry of the boundary curve.} 

\end{table}

We shall now apply a series of AdS symmetries in order to transform this configuration, from the viewpoint of a Poincar\'e observer,  into a Polyakov loop at zero temperature. Let us define a rotation and a boost in a given plane simply by multiplying the coordinate vector of that plane by the matrices
\begin{equation}
\label{review:rot.boost}
	\begin{pmatrix} \cos\alpha & -\sin\alpha \\ \sin\alpha & \cos\alpha \end{pmatrix}~,\qquad
	\begin{pmatrix} \sqrt{1+\gamma^2} & -\gamma \\ -\gamma & \sqrt{1+\gamma^2} \end{pmatrix}~,
\end{equation}
where $\alpha$ and $\gamma$ are the rotation angle and the boost parameter, respectively. Which of the two operations can be used depends, of course, on the metric on the plane. After such a transformation is done, we use \eqref{review:ads.P.coords} with the new $y$-coordinates to obtain the transformed configuration in the Poincar\'e frame. On the boundary, where $\rho=l$, the effect is that of a conformal transformation. For example, in order to obtain a circular Wilson loop of arbitrary radius $R$, one performs a boost in the $(4,5)$ plane with an appropriate $\gamma$. 

\begin{figure}[t]
\begin{center}
\includegraphics[width=0.3\textwidth]{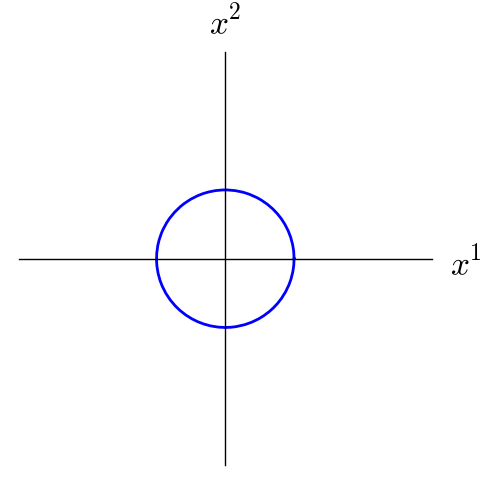}
\includegraphics[width=0.3\textwidth]{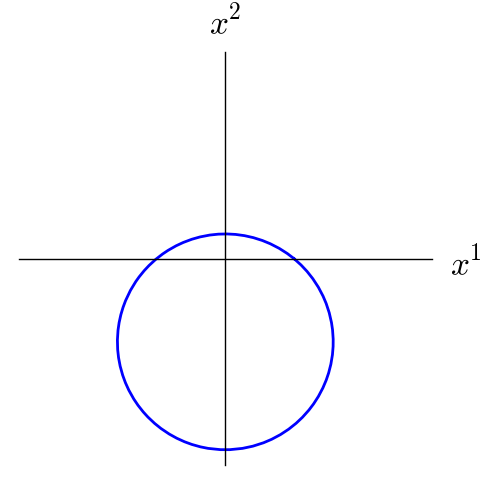}
\includegraphics[width=0.3\textwidth]{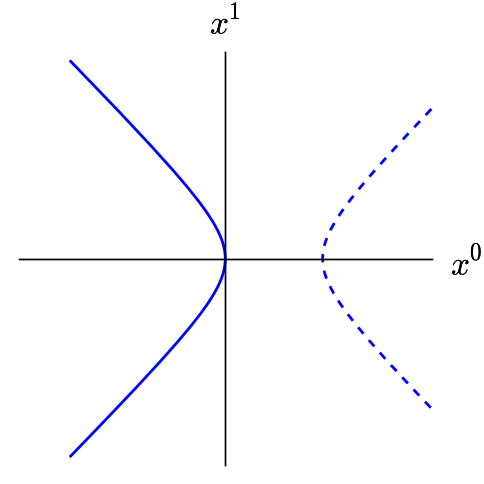}
\end{center}
\caption{\label{review:conf.figs}Boundaries of the configurations \#1, \#2 and \#4 of Table~\ref{review:tab.trans}, respectively. The dashed line indicates that this portion of the global boundary is behind the horizon ($z<0$).}
\end{figure}

We list in Table~\ref{review:tab.trans} the series of transformations that connect the Wilson and Polyakov loops. For some of these configurations, the shape of the boundary curves is illustrated in Figure~\ref{review:conf.figs}.
A few comments on these transformations are in order. One might naively have interpreted the straight line configuration \#3 as the Polyakov loop. However, it is easy to see that the string world-sheet in this configuration never crosses the horizon. The Polyakov loop at finite temperature, in contrast, intersects the horizon on a null line, $\rho=\rh$ for any $\tau$, and one expects this property to survive at zero temperature. Hence, in order to obtain a Polyakov loop, we must perform a transformation that moves part of the world-sheet behind the horizon. This is achieved by a rotation in the $(0,5)$ plane. Combining such a rotation by some angle $\alpha$ with a $(0,4)$ boost with $\gamma=-\cot \alpha$ gives rise to a world-sheet, whose portion with $\sin\tau<0$ is hidden behind the horizon. The particular value $\alpha=\pi/4$ in configuration \#4 has been chosen such that the visible branch of the boundary hyperbola intersects the origin. The final (infinite) boost deforms this branch of the hyperbola into a straight line, while moving the invisible branch to infinity. Finally, doing a Wick rotation of $x_0$ transforms AdS$_5$ into the two separate copies of EAdS$_5$ (see Fig.~\ref{AdS.fig}), with half of the string world-sheet in each copy. These are just two copies of the Polyakov loop. Hence, our symmetry argument confirms what has been found by the explicit calculations.  

